\documentclass[aps,pra,twocolumn,superscriptaddress,amsmath,amssymb]{revtex4-1}
\usepackage[colorlinks,pdfusetitle,urlcolor=blue,citecolor=blue,linkcolor=blue,bookmarksnumbered,plainpages=false]{hyperref}
\usepackage{graphicx}

\usepackage{acro} 
\usepackage{braket}
\usepackage{bm}
\usepackage{amsmath}
\usepackage{siunitx}

\usepackage{tikz}
\usetikzlibrary{arrows, positioning}

\newcommand{\f}[1]{\underline{\bm{#1}}}

\begin{document}

\title{Shape optimizations for body-assisted light--matter interactions}
\author{Jonas Matuszak}
\affiliation{Physikalisches Institut, Albert-Ludwigs-Universit\"at Freiburg,\\
  Hermann-Herder-Str. 3, D-79104 Freiburg i. Br., Germany}
\author{Stefan Yoshi Buhmann}
\affiliation{Institut f\"ur Physik, Universit\"at Kassel, Heinrich-Plett-Stra{\ss}e 40, 34132
Kassel, Germany}
\author{Robert Bennett}
\affiliation{School of Physics \& Astronomy, University of Glasgow, Glasgow,
  G12 8QQ, United Kingdom}
\date{\today}

\begin{abstract}
  We implement a shape optimization algorithm for body-assisted light--matter
  interactions described by the formalism of macroscopic quantum electrodynamics. The
  approach uses the level-set method to represent and incrementally evolve dielectric
  environments. Utilizing finite-difference time-domain techniques we demonstrate the
  ability of the algorithm by optimizing the rate of resonance energy transfer in two
  dimensions. The resulting geometries enhance the transfer rate by several orders of
  magnitude.
\end{abstract}

\maketitle

\section{Introduction}
\label{sec:introduction}

The discovery of photonic and optical designs is usually governed by the symmetry
considerations and experience of the engineer. Traditionally one designs a device and then
investigates its desired properties, here referred to as ``forward'' design. Inverse
design takes the opposite approach by specifying the desired properties and then
algorithmically determining the designing of the final device. Some of the first
applications of optimization algorithms in photonics were genetic algorithms used to
minimize the loss in to fiber-to-ridge waveguide connections \cite{Spueler1998} and
gradient-based algorithms to improve parameter settings in order to maximize band gaps
\cite{Cox1999}. With increasing computational power and the development of the adjoint
methods in aerodynamics \cite{Jameson1988}, the application of inverse design algorithms
now covers a wide variety of areas. In photonics they range from second harmonic
generation \cite{Lin2016} to on-chip wavelength demultiplexers \cite{Piggott2015} or the
optimization of solar energy conversion \cite{Alaeian2012}. A detailed review of the
application in nanophotonics can be found in Ref.~\cite{Molesky_2018}.

In contrast to forward design, where the geometry and materials of a device uniquely
define its electromagnetic properties, the specified properties of the inverse design
problem are not guaranteed to have an existing or unique corresponding design. This leads
to the task of finding a design which is closest to the desired properties, which is done
by maximizing a merit function encoding the desired properties. Inverse design algorithms
usually do not find the global maximum, but converge to a design yielding a local maximum,
which still outperforms most designs found by a forward design approach.

Such techniques from photonics were only recently expanded to virtual-photon-mediated
processes. The authors of Ref.~\cite{Bennett2020} derived a general formulation of the
adjoint optimization methods in the framework of macroscopic quantum electrodynamics. This
allows inverse design techniques to be applied to phenomena such as the Casimir
\cite{Raabe2003} and Casimir-Polder forces \cite{Casimir1948, Buhmann2004}, Van der Waals
forces \cite{Buhmann2004a}, quantum friction \cite{Klatt2017}, resonance energy transfer
(RET) \cite{Foerster1948}, and more. Two alternative ways of approaching the inverse
design algorithm are offered by the formalism. The additive approach taken in
Ref.~\cite{Bennett2020} consists of determining the best possible position of where to add
a small amount of material to the design and thereby iteratively creating an optimized
geometry. The approach taken here is the level-set method \cite{Osher1988}, which consists
of gradually changing the surface of an initial shape towards an
optimal geometry.

This article is structured as follows. First, the general formulation of the optimization
problem in the macroscopic QED framework is discussed and the level-set method is
introduced (Sec.\,\ref{sec:background}). Section \ref{sec:algorithm} illuminates the
implementation of the optimization algorithm and introduces the phenomenon of resonance
energy transfer as an example application. The results of the application are discussed in
Sec.\,\ref{sec:results} and an approach for extending this algorithm to meet manufacturing
constraints is given in Sec.\,\ref{sec:extensions}, which is followed by a summary and
some concluding remarks (Sec.\,\ref{sec:conclusion}).

\section{Background}
\label{sec:background}

In this section we outline the general formalism of inverse design for processes
described within the framework of macroscopic QED. The desired properties of optical
devices are usually represented by a merit function $F$ which depends on the
$\bm{E},\bm{B},\bm{D}$ and $\bm{H}$ fields. In the formalism of macroscopic QED, these
fields are all expressed in terms of the dyadic Green's tensor $\bm{G}$ \cite{Gruner1996}
which is uniquely defined by the inhomogeneous Helmholtz equation
\begin{align}
  \label{eq:helmholtz}
  \Bigg[ \nabla\times \frac{1}{\mu(\bm{r},\omega)} \nabla\times \,\, - 
  \frac{\omega^2}{c^2} \varepsilon(\bm{r}, \omega) \Bigg]
  \bm{G}(\bm{r}, \bm{s}, \omega) = \bm{\delta}(\bm{r}-\bm{s}),
\end{align}
where $\mu (\bm{r}, \omega)$ and $\varepsilon (\bm{r}, \omega)$ are the magnetic
permeability and electric permittivity, respectively. This tensor describes the field
propagation from a source at position $\bm{s}$ to an observation point $\bm{r}$, which
means it encodes information about the electromagnetic environment's geometry as well as
its material response. Noting that $F$ must be an observable we write $F$ as a real-valued
functional of $\bm{G}$:
\begin{align}
  \label{eq:F}
  F = \int\text{d}^3r\int\text{d}^3s\int^\infty_0\text{d}\omega f[\bm{G}(\bm{r},\bm{s},\omega)].
\end{align}
By writing $F$ as an integral we can account for spatially extended sources and multimode
effects as well as for optimization of effects in extended volumes. As derived in more
detail in Ref.~\cite{Bennett2020}, a small addition of material results in a change of the
merit function:
\begin{align}
  \label{eq:delta_F}
  \delta F = & \mu_0\alpha n\int\text{d}^3r \int\text{d}^3s\int_V\text{d}
  r^\prime\int\text{d}\omega \omega^2 \\ \nonumber
  & \times 2\text{ Re }\Bigg\{
  \frac{\partial f} {\partial \bm{G}}(\bm{r},\bm{s},\omega)
    \odot \bm{G}^T(\bm{r}^\prime,\bm{r},\omega)\cdot\bm{G}(\bm{r}^\prime,\bm{s},\omega) \Bigg\},
\end{align}
where $\odot$ represents the Frobenius product $\bm{A}\odot\bm{B} =\sum_{i,j} A_{ij}B_{ij}$
This expression allows one to write $\delta F$ for a material addition at position
$\bm{r}^\prime$ entirely in terms of Green's tensors with sources at $\bm{r}$ and $\bm{s}$
(but not $\bm{r}^\prime$). Calculating the Green's tensors for those two sources is
sufficient to know $\delta F$ at every point $\bm{r}^\prime$.

\begin{figure}
  \centering
  \includegraphics[width=0.25\textwidth]{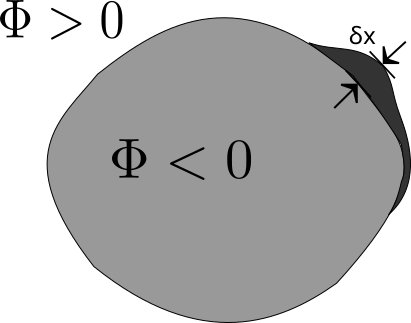}
  \caption{Level-set method. The shape is deformed by a boundary movement $\delta x$.}
  \label{fig:levelset}
\end{figure}

The optimization of optical processes can be approached in several ways. One way is to
calculate $\delta F$ over the region of optimization and simply place additional material
where the change is highest, which is called the additive approach. Repeating the
calculation of $\delta F$ and placing the new material at the new position of highest
change gives the simple iterative optimization algorithm used in Ref.~\cite{Bennett2020}.
In this work we take a different approach, the level-set method. We start with an initial
geometry and gradually change its boundaries, as sketched in Fig.~\ref{fig:levelset}.

The geometry of two different materials can be conveniently represented by a level-set
function $\Phi$ \cite{Osher1988}. The boundary is represented by $\Phi = 0$. Inside the
boundary, the level-set function takes negative values and outside it takes positive
values. The initial shape of the boundaries is described by
\begin{align}
  \Phi(\bm{r}(t), t) = 0, \label{eq:phi_init}
\end{align}
where the artificial `time' parameter $t$ describing the iterative process is introduced.
This parameter is used to keep track of the boundaries and their deformation. The total time
derivative of (\ref{eq:phi_init}) yields
\begin{align}
  \frac{\partial \Phi}{\partial t} + \nabla\Phi \cdot \frac{\partial \bm{r}}{\partial t} = 0,
\end{align}
which is the advection equation well-known from fluid dynamics, describing transport of a substance
in a velocity field. 
Since only movement orthogonal to the boundary is relevant, the equation can be simplified by
introducing the scalar velocity field in the normal direction of the boundary $v_n$:
\begin{align}
  \label{eq:advection}
  \frac{\partial \Phi}{\partial t} + v_n |\nabla\Phi| = 0.
\end{align}
If we now want to evolve the boundaries of the geometry in such a way that $F$ increases,
we have to choose an appropriate velocity field. For this we can use the information
gained from $\delta F$. Where the change in $F$ is positive at a point close to (but
outside) the boundary, the velocity should be positive, leading the boundary to deform in
such a way that this point becomes included within. To ensure this we can rewrite the
integral in (\ref{eq:delta_F}) over the volume as
\begin{align}
  \int_V\text{d}^3r^\prime \rightarrow \int_{\partial V}\text{d}A\delta x(\bm{r}^\prime)
  = \int_{\partial V} \text{d}A v_n\delta t.
\end{align}
The infinitesimal change $\delta x$ of the boundary is replaced by the product of an
infinitesimal time step $\delta t$ and the velocity normal to the boundary $v_n$.
By choosing the velocity field to be
\begin{align} \label{eq:v_n}
  v_n = 2 \text{ Re }& \alpha n\mu_0 \int\text{d}^3r\int\text{d}^3s \int\text{d}\omega
  \omega^2 \\ \nonumber
   &\times\frac{\partial f} {\partial \bm{G}}(\bm{r},\bm{s},\omega)
  \odot \bm{G}^T(\bm{r}^\prime,\bm{r},\omega)\cdot\bm{G}(\bm{r}^\prime,\bm{s},\omega),
\end{align}
we ensure a positive change of the merit function for small time steps $\delta t$:
\begin{align}
  \delta F  = \int_{\partial V}\text{d}A v_n^2 \delta t. \label{eq:delta_F_v_n}
\end{align}
With this we have arrived at a general expression to calculate changes in merit functions
which are expressible in terms of the Green's tensor.

Before proceeding, we caution about one complication that has not been
taken into account in the above analysis.
While the parallel component of $\bm{G}$ is continuous across an interface, continuity
of the product $\varepsilon \bm{G}$ holds for the perpendicular component
(in exactly the same way as the perpendicular component of electric displacement $\bm{D}$
is continuous across an interface). This could lead to a situation where using a large
value of $\varepsilon$ causes a point which was outside the interface before optimization to have a
significantly different $\bm{G}$ if the boundary deforms so as to include it. This would
entail taking higher-order terms in the Born series that leads to Eq.~(\ref{eq:delta_F})
to ensure convergence \cite{PhysRevE}. In this work we simply note that the atoms
themselves are far enough away from any surface for this problem to have any effect, so
while there are imperfections in the optimization algorithm the RET rate calculated for
the resulting structures is reliable.

\section{Level-Set Algorithm}
\label{sec:algorithm}

We implement the level-set method as indicated by Fig.~\ref{fig:flow_chart}.
One starts by choosing an initial geometry as an input to the algorithm, which one encodes
in the level-set function $\Phi$. Usually $\Phi$ is chosen to be the signed distance function
from the shape boundaries.

Next the Green's tensor has to be computed. As seen in Eqs.\,(\ref{eq:v_n}) and
(\ref{eq:delta_F_v_n}) computing $\bm{G}$ for a source located at both $\bm{s}$ and
$\bm{r}$ is sufficient to know the change in the merit function for a material addition at
any point $\bm{r}^\prime$ in the optimization domain. With the aid of the Green's tensors, the
velocity field $v_n$ is obtained and used in the next step to evolve the boundaries of the
level set function by solving the advection equation (\ref{eq:advection}) for a short time
step $t$. This results in a small deformation of the original shape leading to an increase
in $F$. The next step is to update the material geometry in the simulation according to
the evolved level set function. Here the process repeats and the best shape deformation is
calculated for the new geometry. This process can be terminated after a certain amount of
iterations or when the merit function no longer increases.

\tikzstyle{block} = [rectangle, draw, fill=gray!30, 
text width=10em, text centered, minimum height=3em, line width=1.6]
\tikzstyle{arrow} = [thick,->, >=stealth, line width=1.5]

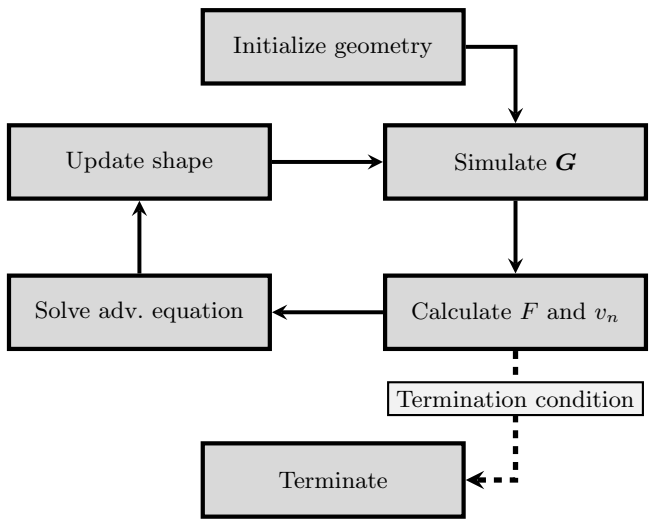
\begin{figure}
\begin{tikzpicture}[node distance = 2cm, auto]
    \node (0,0) [block] (init) {Initialize geometry};
    \node [block, below right=0.5cm and -1.1cm of init] (sim) {Simulate $\bm{G}$};
    \node [block, below of=sim] (calc) {Calculate $F$ and $v_n$};
    \node [block, left of=calc, node distance=5cm] (solve) {Solve adv.~equation};
    \node [block, left of=sim, node distance=5cm] (update) {Update shape};
    \node [block, below left=1.2cm and -1.1cm of calc] (term) {Terminate};
    \draw [arrow] (init) -| (sim);
    \draw [arrow] (sim) -- (calc);
    \draw [arrow] (calc) -- (solve); 
    \draw [arrow] (solve) -- (update);
    \draw [arrow] (update) -- (sim);
    \draw [dashed, ->, >=stealth,line width=2] (calc) |- node[above=0.85cm, solid, thick,
    rectangle, draw, fill=gray!10] {Termination condition} (term);
  \end{tikzpicture}
  \caption{Flow chart of the inverse design algorithm. It starts with an initial geometry and
    then calculates $\bm{G}$ from simulations with MEEP. Using $\bm{G}$, the merit function and
    the velocity field $v_n$ are obtained and the latter is used in the advection equation, which
    is solved for a short time step $t$. With the new geometry a simulation is started and
    the process repeats. Under a condition, e.g. a vanishing increase in $F$, the process
    is terminated.}
  \label{fig:flow_chart}
\end{figure}

\subsection{Computational Approach}

The Green's tensor is known in analytical form only for a few highly symmetric geometries.
Here we encounter arbitrary shapes, which is why we use a numerical approach. We use
existing electrodynamics simulation software, here the open source package MEEP
\cite{meep}, which uses finite-difference time-domain methods, and extract information
about $\bm{G}$ from the electric field.

As discussed in \cite{Bennett2020}, we can relate the Fourier transform of the $E$-field
directly to the Green's tensor and a source current $\f{j}$. For a point current source
$\f{j}(\bm{r}^\prime,\omega) = \delta(\bm{r}^\prime-\bm{s})\f{j}(\omega)$, we have
\begin{align}
  \f{E}(\bm{r},\omega) = \text{i}\mu_0\omega \bm{G}(\bm{r},\bm{s}, \omega)\cdot\f{j}(\omega).
\end{align}
By rearranging, a given component of $\bm{G}$ can be calculated with the simulated fields
and the Fourier transformed point current source:
\begin{align}
  G_{ij}(\bm{r},\bm{s},\omega) = \frac{\underline{E}_i(\bm{r},\omega)}{\text{i}\mu_0\omega
  \underline{j}_j(\omega)} \label{eq:G_conv}.
\end{align}
MEEP provides a built-in Gaussian-shaped source current,
\begin{align}
  \label{eq:gaussian}
  \bm{j}(t) = \bm{j}_0\exp\Big\{-2\pi \text{i} f t - \frac{(t- t_0)^2}{2w^2} \Big\},
\end{align}
with frequency $f$, time of maximal amplitude $t_0$ and width $w$.
For the simulation the time of maximal amplitude is set to be five times the
temporal width of the Gaussian. Furthermore, a cutoff to the fields is applied after the
Gaussian has decayed for five widths.

The advection equation (\ref{eq:advection}) must also be solved numerically. For this we
use the finite volume solver for partial differential equations FiPy \cite{FiPy}, which
incorporates a second-order upwind scheme.

\subsection{Resonance energy transfer}

Here we apply the algorithm as described above to the process of resonance energy transfer
\cite{Ho2002,Hemmerich2018}. RET consists of two atoms interacting through virtual photon
exchange. An initially excited donor atom emits a photon which is absorbed by the acceptor
atom in the ground state. This can be well approximated as a single-frequency phenomenon
where we also use the dipole approximation to model the donor and acceptor. In order to
apply this algorithm to RET we have to express the change in the transfer rate by the
means of a merit function. For an acceptor and donor dipole moment $\bm{d}_\mathrm{A}$ and
$\bm{d}_\mathrm{D}$ the expression is well known as
\begin{align}
  \Gamma = \frac{2\pi \mu_0^2\omega^4_\mathrm{D}}{\hbar}  |\bm{d}_\mathrm{A}^*\cdot
  \bm{G}(\bm{r}_\mathrm{A}, \bm{r}_\mathrm{D}, \omega_\mathrm{D}) \cdot \bm{d}_\mathrm{D}|^2.
\end{align}
This is already expressed in terms of the Green's tensor, so we can assign the merit function
$F$ to be equal to $\Gamma$. 
It is easy to see that the choice 
\begin{align}
  f = \frac{2\pi \mu_0^2\omega^4}{\hbar} &|\bm{d}_\mathrm{A}^*\cdot \bm{G}(\bm{r},
  \bm{s}, \omega) \cdot \bm{d}_\mathrm{D}|^2 \\ \nonumber
  & \times \delta(\omega-\omega_\mathrm{D})\delta(\bm{r}-\bm{r}_\mathrm{A})
  \delta\bm(\bm{s}-\bm{r}_\mathrm{D})
\end{align}
reproduces the merit function $F = \Gamma$. Using $f$, we calculate the velocity field
which increases $F$ for a small deformation of the shape. According to Eq.~(\ref{eq:v_n})
and with the use of some algebra, $v_n$ becomes
\begin{align} \label{eq:v_ret}
  v_n(\bm{r}^\prime) &= \frac{4\pi\alpha n\mu_0^3\omega^6_\mathrm{D}}{\hbar} \text{ Re } \Bigg\{
  \bm{d}_\mathrm{A}\cdot\bm{G}^*(\bm{r}_\mathrm{A},\bm{r}_\mathrm{D},\omega)\cdot\bm{d}_\mathrm{D}^* \nonumber\\ 
  & \times \Big[ \bm{d}_\mathrm{A}^*\cdot\bm{G}^T(\bm{r}^\prime,\bm{r}_\mathrm{A},\omega)\Big]\cdot
    \Big[\bm{G}(\bm{r}^\prime,\bm{r}_\mathrm{D},\omega)\cdot\bm{d}_\mathrm{D}\Big]\Bigg\}.
\end{align}

\section{Results}
\label{sec:results}

In order to illustrate our method, we restrict the application to systems with
translational invariance along one axis, reducing the computational effort to a two
dimensional problem. We apply the algorithm to two-dimensional (2D) RET with an initial
material distribution shaped as a cylinder. In the following we quantify the optimization
by a dimensionless ratio $Q$,
\begin{align}
  Q = \frac{\Gamma}{\Gamma_0},
\end{align}
comparing the RET rate $\Gamma$ of a geometry to the free space rate $\Gamma_0$, broadly analogous
to the Purcell factor of spontaneous emission.

The transition wavelength is set to \SI{2}{\micro\metre}, the dipoles are aligned along
the $x$-axis and the simulation is run for 300 iterations. The convergence of the
algorithm is discussed in the Appendix. In Fig.~\ref{fig:cylinder}, the
results of the optimization process are shown at four iteration steps ranging from the
initial shape to the iteration of highest optimization. The highest optimization is
reached with a Purcell factor of approximately $Q = 7\cdot 10^5$ at the $278^\mathrm{th}$
iteration.
Within the first iterations the shape grew mainly in its width towards the dipoles and
later developed ``armlike'' structures around the dipoles, since the impact of matter
placed is higher for small distances to the dipoles. In the direction of dipole alignment
the arm structures are open, which can be explained by much weaker radiation in this
direction compared to the perpendicular direction. After the development of the first arm
structure, a second pair of arms evolved, with a distance of approximately
\SI{1}{\micro\metre} at the far end of the shape towards \SI{0.5}{\micro\metre} at their
origin, which corresponds to half and a quarter of the transition wavelength respectively.
This double arm structure is reminiscent of a waveguide, focusing the electromagnetic
radiation from the donor dipole to the acceptor dipole.

\begin{figure}
  \centering
  \includegraphics[width=0.5\textwidth]{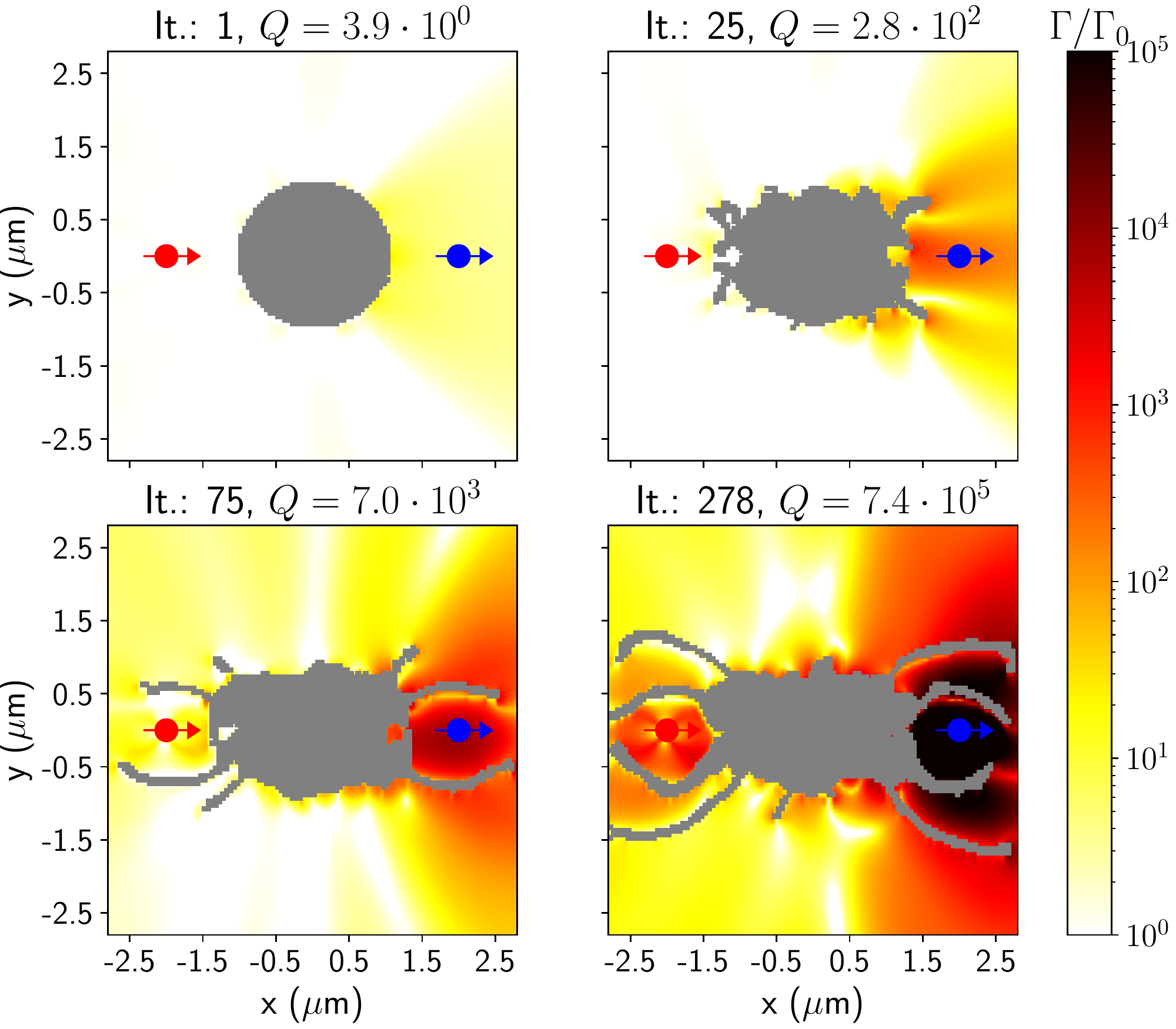}
  \caption{Inverse design process for 2D RET with the initial shape of a cylinder. The
    upper left panel shows the initial shape, the red dot is the donor dipole, the blue
    dot is the acceptor. The arrow through each dot shows the dipole orientation; here
    they are aligned in the $x$ direction. The simulations are carried out with a
    resolution of 20 pixels per \si{\micro\metre} and the time interval $t$ of the
    evolution of $\Phi$ is chosen in such a way, that the boundary moves maximally 0.1
    \si{\micro\metre} (step size) per iteration. The evolution of the shape is shown by
    the iteration steps 25 and 75 and the step of highest optimization, with
    $Q \approx 7\cdot 10^5$.}
  \label{fig:cylinder}
\end{figure}
\begin{figure}
  \centering
  \includegraphics[width=0.5\textwidth]{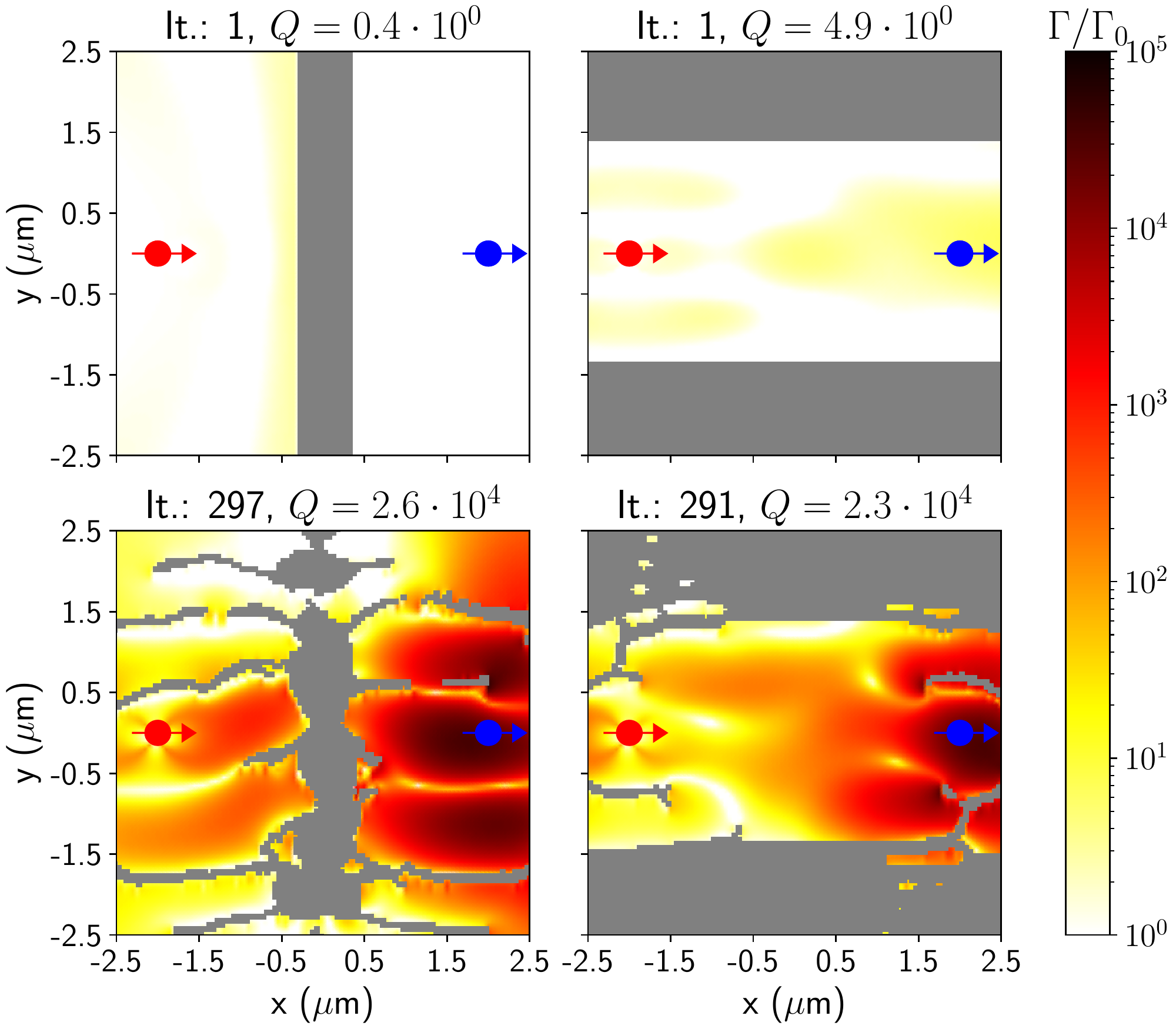}
  \caption{Results of the optimization for different initial shapes. The first column
    shows the result of starting with a vertical wall between the dipoles, the second
    shows the result for a waveguide like structure. The first row shows the initial
    shapes, the second shows the iteration of the highest amplification for
    the processes.}
  \label{fig:other_shapes}
\end{figure}

Next we apply the inverse design algorithm with the same settings to the initial shapes of
a vertical wall and a waveguide (Fig.~\ref{fig:other_shapes}). The optimization for the
shapes is of the same order of magnitude with $Q \approx 10^4$. As with the cylinder, the
initial wall is modified by the appearance of arm structures, but due to the larger
distance of the dipoles to the material, the algorithm lets the shape first develop
waveguide structures above and below the dipoles which are parallel to the dipole
orientation and with a width of half the transition wavelength.

The initial waveguide shape also develops arm structures along the dipoles, but in
contrast to the other shapes the algorithm places no material between the dipoles. This
can be explained by the dipole orientation along the $x$-axis. The dipole radiation in
this direction is weak compared with radiation perpendicular to the orientation, which
also results in a stronger deformation of the initial structure above and below the
dipoles. Above the donor dipoles small cavities are visible which are periodic in half the
transition wavelength in material
$\lambda^\prime = \lambda/\sqrt{\varepsilon} \approx \SI{0.58}{\micro\metre}$. We interpret these as having the same
functionality as a Bragg mirror, reflecting radiation efficiently back into the initial
waveguide structure.


We also want compare the shape optimization method to the additive method
used in \cite{Bennett2020}. With the additive method the authors achieved a optimization
of $Q \approx 10^{5}$ within 250 iteration steps. The shape optimization algorithm achieved
similar optimization of $Q \approx 7\cdot10^{5}$ after 278 iteration steps as seen in
Fig.~\ref{fig:cylinder}, while other shapes stay an order of magnitude below that
(Fig.~\ref{fig:other_shapes}). So, depending on the choice of the initial shape, the
performance of the two methods is quite similar.
  
The main qualitative features of the emerging structures are also comparable. For both
methods we see thin armlike structures evolving (see Fig.~5 in \cite{Bennett2020}).
However, the two algorithms also lead to intrinsic structural differences. The additive
method places new material at the point of highest impact, which results in a more
material--efficient optimization. However the shapes created are also disconnected which
may not be feasible to manufacture -- with the shape optimization approach we mostly end up
with continuous structures. With the shape optimization approach it is also possible to
impose manufacturing constraints, since one has a direct representation of the shape in
form of the level-set function on which one can place restrictions. One possible
constraint will be discussed in Sec.~\ref{sec:extensions}. The level-set method adds the
computational cost of solving the advection equation, however compared to the calculation
of the Green's tensor it only adds 5\% of additional computation time in our
implementation.

\subsection{Rotating Dipoles}

\begin{figure}[t]
  \centering
  \includegraphics[width=0.5\textwidth]{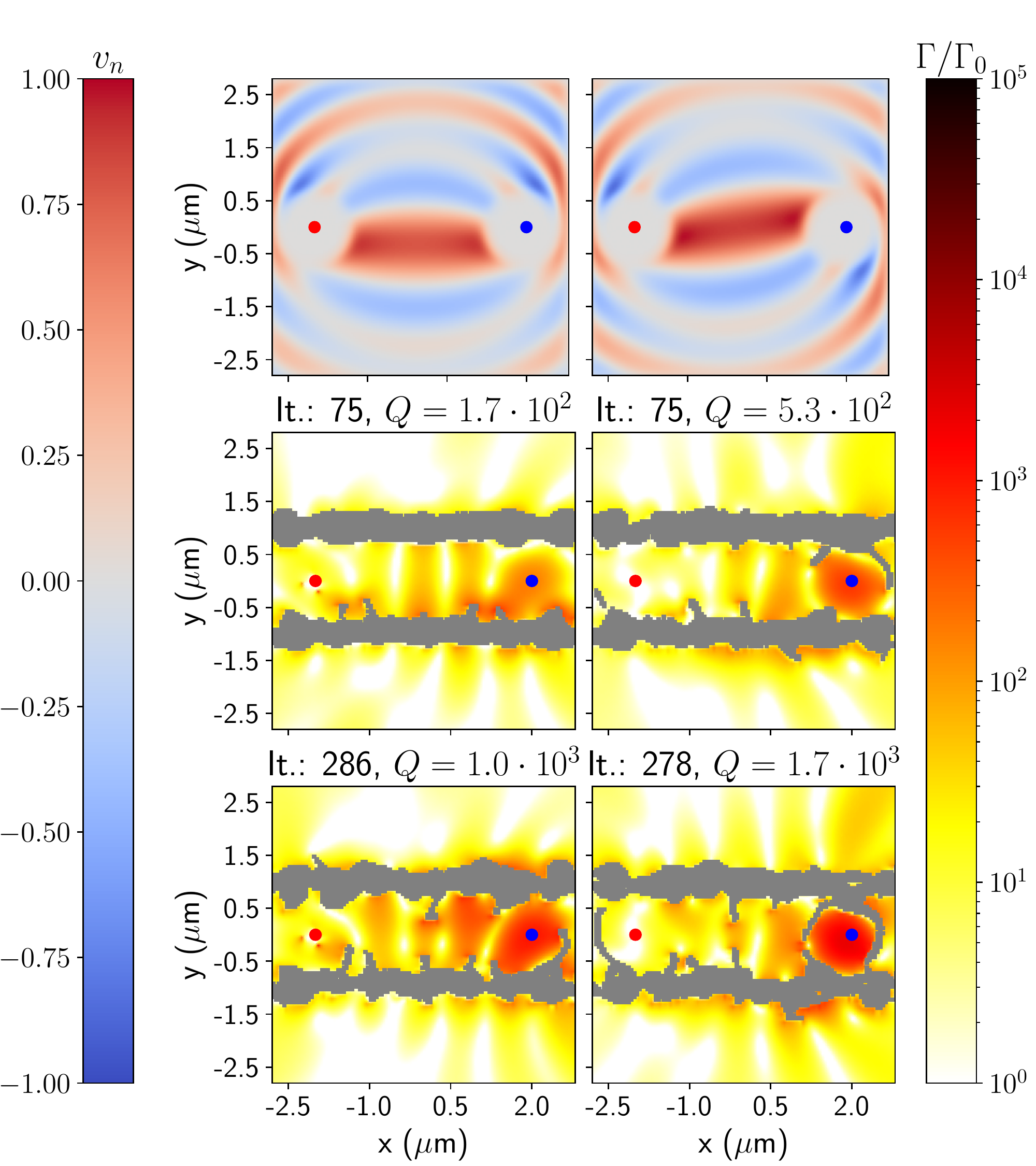}
  \caption{ Optimization results for co- (left) and counter-rotating (right) dipoles for
    the initial shape of two bars above and below the dipoles. The simulations were run
    with a resolution of 20 pixels per \si{\micro\metre}, a transition wavelength of
    \SI{2}{\micro\metre} and a permittivity of the material of $\varepsilon$ = 12. In the upper
    panels $v_n$ is shown for free space for co- and counter-rotating dipoles
    respectively. The velocity fields show asymmetry along the $x$-axis for co-rotating
    and asymmetry along the $x$- and $y$-axis for the counter-rotating dipole. The two
    white circles are the positions of the dipoles where the velocity was set to zero by
    hand. The second row shows the shapes after 75 iterations, the third row shows the
    shapes shows the shapes for the iteration of highest optimization, which is of the
    order of $10^3$ for both dipole rotations.}
  \label{fig:rotating}
\end{figure}

The case of rotating dipoles has been brought into focus by recent predictions of effects
such as lateral interatomic forces and asymmetric emission
\cite{Barcellona2020,Ref20,Ref21,Ref26}. In order to observe such phenomena, it may be
useful to consider optimized environments for them. For this reason, we apply the
algorithm to co- and counter-rotating dipoles. Similarly to circular polarized light we
describe them as complex valued vectors. A dipole rotating in the $xy$ plane becomes
$\bm{d} = (1,\text{i},0)^\text{T}$ and a dipole counter-rotating with respect to $\bm{d}$
reads $\bm{d}^\prime = (1, -\text{i}, 0)^\text{T}$. Fig.~\ref{fig:rotating} shows the velocity
field for the co- and counter-rotating dipoles in the case of free space and the 75th
iteration as well as the iteration with the highest $Q$. The velocity fields are now no
longer symmetric along the $x$-axis and the spiral velocity field of the counter-rotating
also does not show symmetry along the $y$-axis. The highest increase for matter placement
in the free space case for the co-rotating dipoles would be below the $x$-axis, whereas
the increase for the counter-rotating dipoles is rotated around the coordinate origin.

These changes are reflected in the evolving shapes. The shape of the co-rotating dipoles
develops structures faster at the lower bar, while the shape of the counter-rotating
dipoles encloses the donor dipole from below and the acceptor dipole from above, following
the distribution of increase in the free space velocity field. The optimization of the
magnitude of $10^{3}$ for both rotations is considerably lower than for the shapes with
linearly polarized dipoles (which achieved a order of magnitude higher optimizations).

\section{Algorithm Extensions}
\label{sec:extensions}

The algorithm can be extended in order to fulfill manufacturing constraints on the material
or shapes. We demonstrate this by constraining the curvature of the evolving shape.

In Ref.~\cite{Piggott2017} a technique was introduced where the level-set function is
evolved separately with a velocity field proportional to its curvature:
\begin{align}
  \kappa = \nabla\cdot \Bigg(\frac{\nabla \Phi}{|\nabla \Phi|}\Bigg).
\end{align}
A weighting function was introduced in order to specify a maximal curvature $\kappa_0$
below which the velocity field is set to zero.
\begin{align}
  b(\kappa) =
  \begin{cases}
    \kappa &\text{ for } |\kappa| > \kappa_0, \\
    0      &\text{ otherwise. }
  \end{cases}
\end{align}
By evolving $\Phi$ with the velocity field $v_\kappa = -b(\kappa)\kappa$ to its steady state,
all features with a curvature above $\kappa_0$ are eliminated.

Our approach is similar; however, we use a localized velocity field $v_\Gamma$ which only
acts in the neighborhood of the surface instead of on the whole level--set function.
We choose $v_\Gamma$ to be proportional and opposite to the curvature of $\Phi$
\begin{align}
  v_\Gamma = -\tau \kappa G(\bm{d}),
\end{align}
where $\tau$ is a factor to scale the velocity field and a Gaussian is used to 
localize this velocity on the surface:
\begin{align}
  G(\bm{d}) = \exp\Bigg\{\frac{|\bm{d}|^2}{\sigma}\Bigg\}. 
\end{align}
Here $\bm{d}$ is the shortest distance to the surface and $\sigma$ is set to a value, such
that the Gaussian's width covers a few pixels. Now the advection equation is solved for
the velocity $v_t = v_n + v_\Gamma$.\\
\begin{figure}
  \centering
  \includegraphics[width=0.5\textwidth]{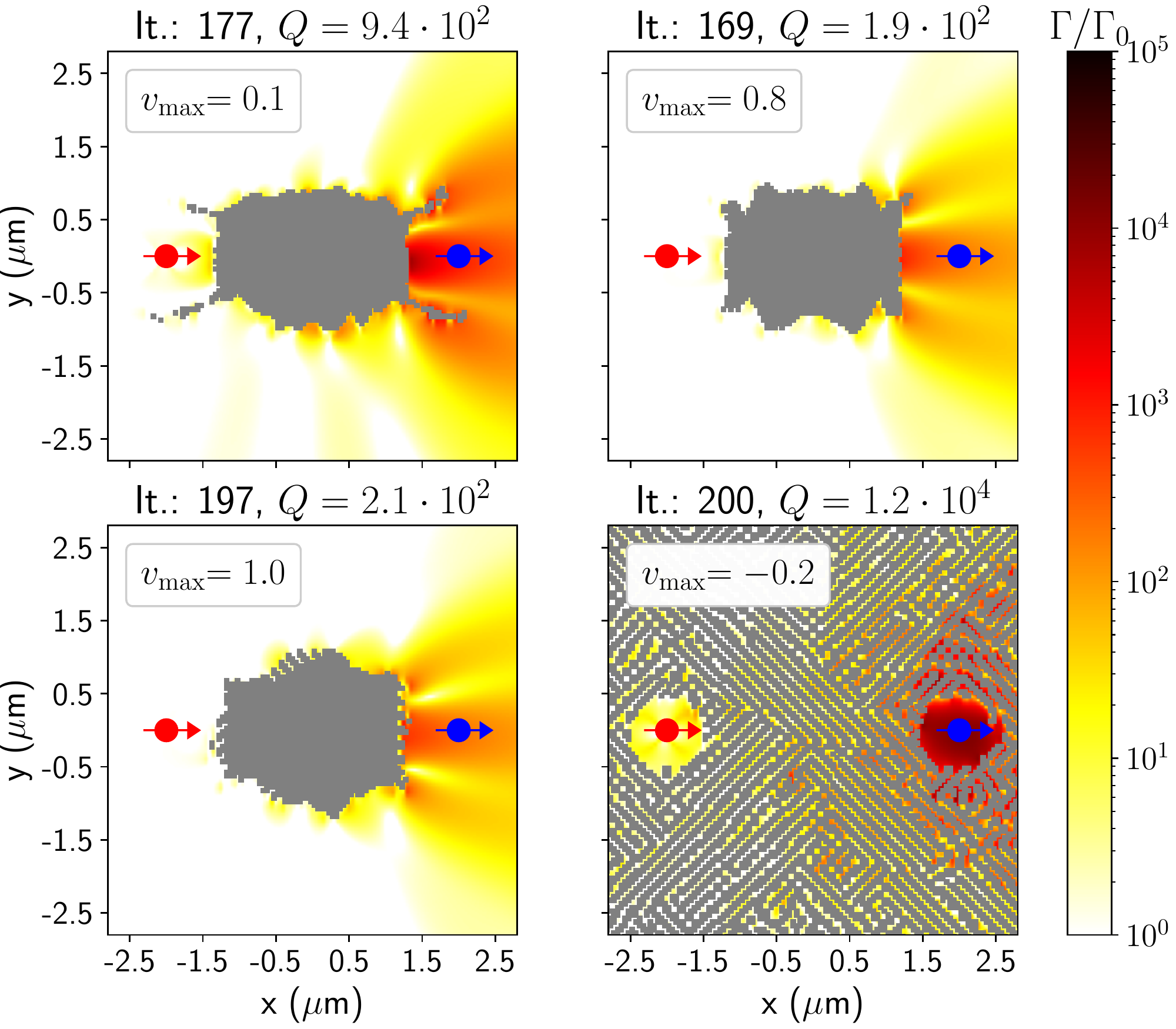}
  \caption{Results of the optimization algorithm with the initial shape of an cylinder and
    surface tension term in the velocity field. The velocity contribution from the surface
    tension is scaled so that its maximal value ranges from 0.1 to 0.8 to 1.0 in the first 3
    plots. In the lower right plot the contribution is inverted, such that the curvature
    is enhanced.}
  \label{fig:surfacetension}
\end{figure}
In Fig.~\ref{fig:surfacetension} the results of this extension applied to the initial
shape of a cylinder are shown. We have normalized the additional velocity field in such a
way, that the maximal value amounted to $v_{\Gamma, max} = 0.1, 0.8, 1$ and
$v_{\Gamma, max}=-0.2$. For the last value we inverted the direction of the velocity field to
see the effects of enhancing the curvature of the shape. It can be observed that the thin
structures yielding a high contribution to the surface decrease with increasing
$v_{\Gamma, max}$, whereas for the negative $v_{\Gamma, max}$ the surface is being maximized by the
emergence of hole structures. Generally we can see that the optimization is of the order
of $10^2$ which is 3 orders of magnitude lower than for the freely evolving shapes.

\section{Conclusion}
\label{sec:conclusion}

In this work we have implemented an efficient shape optimization algorithm using the
level--set and adjoint methods. The example application to two dimensional resonance
energy transfer for different initial structures has yielded shapes with optimizations of
several orders of magnitude in the transfer rate. We also demonstrated an example of how
the algorithm can be extended to meet manufacturing constraints by restricting the
curvature of the optimized shape. Since the method is based on the dyadic Green's tensor
it can be applied to a vast variety of other processes and interactions characterizeable
by this tensor. Further work could include three dimensional shape optimizations,
applications to other processes and implementations of different manufacturing
constraints.

\appendix

\section{Algorithm Convergence}
\label{sec:convergence}
We briefly investigate the convergence of the algorithm towards a locally optimal
design. The accuracy of the simulations with MEEP are mainly determined by their
resolution, which is also true for the solving of the advection equation. A second
influence on the algorithm is the volume of material added per iteration.

\subsection{Resolution Dependence}
\label{sec:res_dependence}

\begin{figure}
  \centering
  \includegraphics[width=0.45\textwidth]{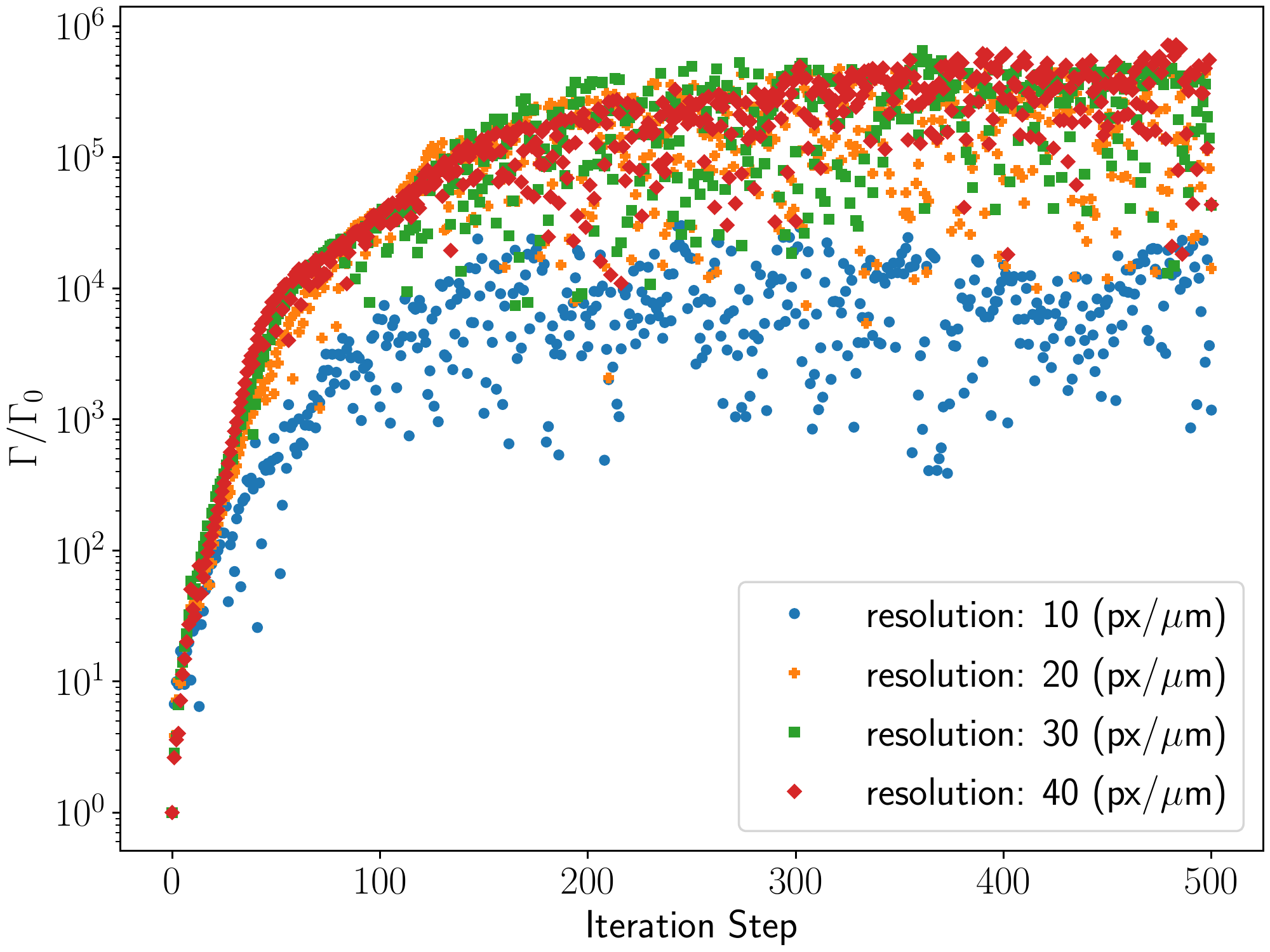}
  \caption{Resolution dependence on the convergence of the Purcell factor $Q$. The
    simulations were performed with an initial shape of a cylinder and a wavelength of
    \SI{2}{\micro\metre}. The resolution is given in pixel per \si{\micro\metre}. The
    inverse design processes perform similarly with the exception of the process with a
    resolution of 10 pixels per \si{\micro\metre}. The oscillations in the Purcell factor
    become prominent in the last 400 iteration steps and increase with lower resolutions.}
  \label{fig:conv_res}
\end{figure}

To test the convergence of the algorithm depending on the simulation resolution we ran
simulations with resolutions ranging from 10 pixels per \si{\micro\metre} up to 40 pixels
per \si{\micro\metre}. As an initial shape we chose a cylinder of radius
$R = \SI{1}{\micro\metre}$, a simulation size of $7\times\SI{7}{\micro\metre}^2$, a
transition wavelength of $\lambda = \SI{2}{\micro\metre}$, the dipole orientations along
the $x$-axis and a maximal change of the border position of \SI{0.1}{\micro\metre} per
iteration. The dipole separation is \SI{4}{\micro\metre} and the material has an
permittivity of $\varepsilon = 12$. The processes were run for 500 iterations for which
the evolution of the Purcell factor $Q$ is shown in Fig.~\ref{fig:conv_res}.

The $Q$ factors for resolutions of 20, 30 and 40 pixels per \si{\micro\metre} show very
similar behavior, they rise four orders of magnitude within the first 100 iterations and
then continue to rise to the order of $Q \approx 10^5 $ within the next 400 iteration
steps. Here the optimization shows oscillations which become more dramatic with lower
resolutions. In the regime of high optimizations the necessary adjustments of the shapes
towards an increased optimization are smaller than in the beginning. The processes with
lower resolutions overstep the best amount of change, generating larger oscillations,
whereas less coarse resolutions allow for a finer adjustment to the velocity field. The
lowest tested resolution of 10 pixels per \si{\micro\metre} rises slower than the other
processes and reaches a plateau at an optimization of $Q \approx 10^4$. With lower
resolution the volume of the pixelwise addition of material increases, which stretches the
limits of the Born approximation, giving another reason for the increase of the
oscillations with lower resolution. This resolution is too low to produce optimal results,
while the resolutions from 20 pixels per \si{\micro\metre} seem suitable for the
simulations.
\subsection{Stepsize Dependence}

\begin{figure}
  \centering
  \includegraphics[width=0.45\textwidth]{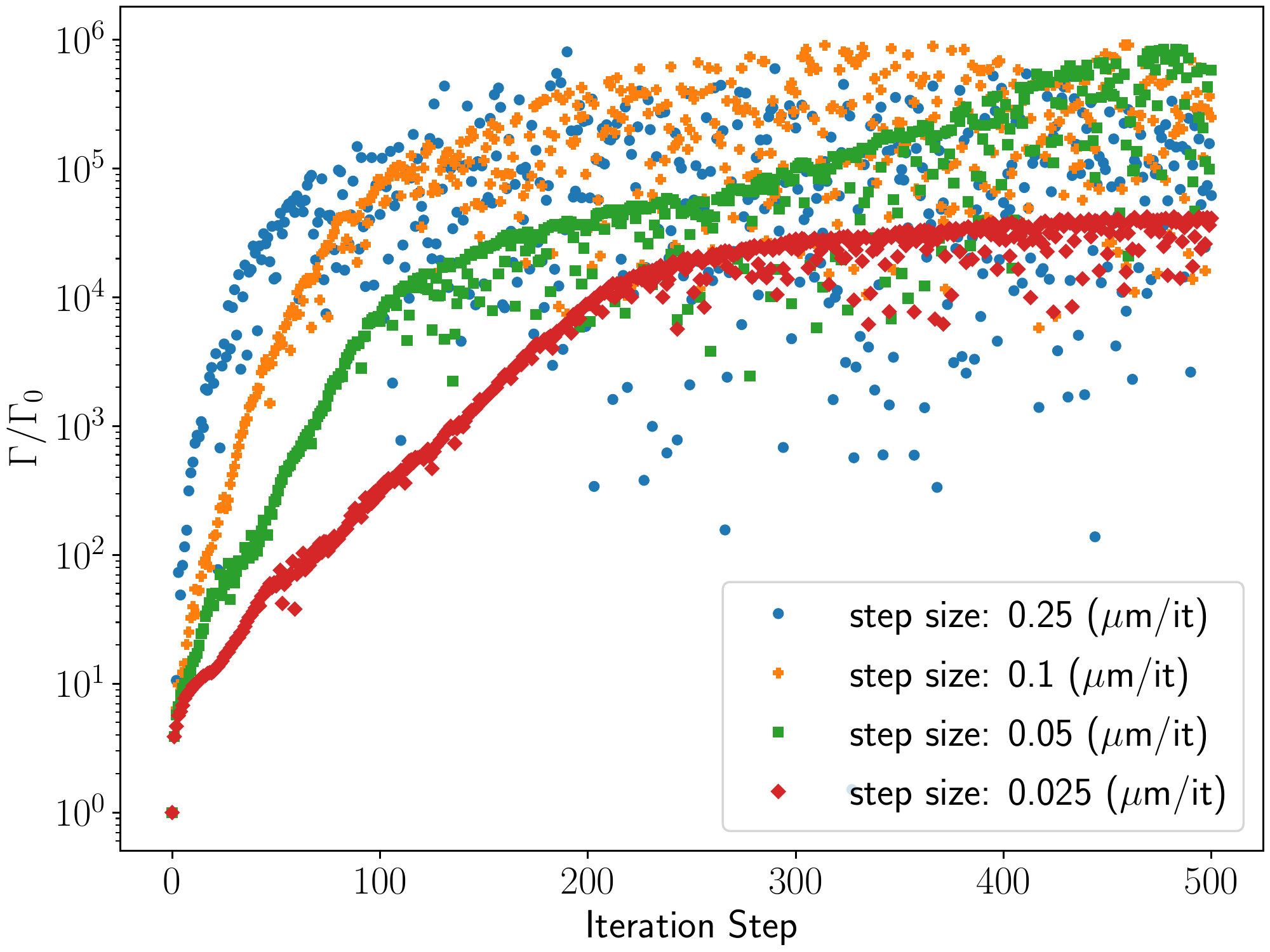}
  \caption{Dependence of $Q$ on the evolution time of the advection equation. A step size of
    0.5 corresponds to an evolution time which lets the boundary move maximally 0.5 pixels
    per iteration. The processes with the presented step sizes all converge to a $Q$ factor
    of the same order of magnitude. Smaller step sizes generally take more iteration steps to
    reach an optimal design but show less oscillations in $Q$. Larger step sizes converge within
    fewer steps, but show large oscillations in $Q$.}
  \label{fig:conv_stepsize}
\end{figure}

Next we investigate the dependence of $Q$ on the magnitude of the boundary deformation,
here referred to as the step size. Since we normalize the velocity field such that its
maximal value is 1, we control the step size via the time for which $\Phi$ is evolved by
the advection equation. We again use the initial cylinder shape and the same parameters as
before. The resolution was set to be 20 pixels per \si{\micro\metre} and optimizations
were performed with a step size between $0.025$ and $0.25$ \si{\micro\metre} per
iteration. The results are shown in Fig.~\ref{fig:conv_stepsize}. Here we see that all
processes reached the same order of magnitude for the enhancement, around $Q = 10^5$,
except for the smallest step size. Similar to the resolution dependence, $Q$ rises for the
three largest step sizes at least four orders of magnitude within the first 100
iterations, while the Purcell factor of the smallest step size reaches this optimization
after 100 more iterations. Since the step size is directly related to the amount of matter
added per iteration, the Purcell factor for optimizations with larger step sizes rises
more quickly. The oscillations which appear after the Purcell factor reaches the order
magnitude of $10^4$ also increase with the step size. As mentioned in the
App.~\ref{sec:res_dependence} this is related to the finer adjustment to the given
velocity field, for the smallest step size these oscillations are considerably smaller and
less frequent. Another reason for the emergence of the oscillation is that we force the
shape through normalization of the velocity field to evolve by the set step size. Here the
algorithm is forced to change the boundaries, even if the optimal change in the boundary
position is smaller than the size of a single pixel.
\\
We conclude that the algorithm reaches higher optimizations faster the larger the step
size, but introduces larger fluctuations in the Purcell factor between iteration steps.
All of the tested step sizes eventually lead to the same order of magnitude of
optimization, except the smallest of \SI{0.025}{\micro\metre}.
%


%

\end{document}